\documentclass[a4paper,11pt]{article}
\pdfoutput=1 

\usepackage{jheppub} 

\usepackage[T1]{fontenc} 
\usepackage[utf8]{inputenc}
\usepackage{amsmath}
\usepackage{float}
\usepackage{lineno}
\usepackage{xcolor}
\usepackage{subfig}
\usepackage{array}

\usepackage[outdir=source/figures/]{epstopdf}
\usepackage{amssymb}
\usepackage{accents}

\emailAdd{vcarrretero@km3net.de}
\emailAdd{km3net-pc@km3net.de}

\abstract{
In the era of precision measurements of neutrino oscillation parameters, it is necessary for experiments to disentangle discrepancies that may indicate physics beyond the Standard Model in the neutrino sector. KM3NeT/ORCA is a water Cherenkov neutrino detector under construction
and anchored at the bottom of the Mediterranean Sea. The detector is designed to study the
oscillations of atmospheric neutrinos and determine the neutrino mass ordering. This paper focuses on the initial configuration of ORCA, referred to as ORCA6, which comprises six out of the foreseen 115 detection units of photosensors. A high-purity neutrino sample was extracted during 2020 and 2021, corresponding to an exposure of 433 kton-years. This sample is analysed following a binned log-likelihood approach to search for invisible neutrino decay, in a three-flavour neutrino oscillation scenario, where the third neutrino mass state $\nu_3$ decays into an invisible state, e.g. a sterile neutrino. The resulting best fit of the invisible neutrino decay parameter is $\alpha_3 = 0.92^{+1.08}_{-0.57}\times 10^{-4}~\mathrm{eV^2}$, corresponding to a scenario with $\theta_{23}$ in the second octant and normal neutrino mass ordering. The results are consistent with the Standard Model, within a 
$2.1\,\sigma$ interval.
}

\begin{document}

\title{Probing invisible neutrino decay with the first six detection units of KM3NeT/ORCA}

\author[a]{S.~Aiello}
\author[b,be]{A.~Albert}
\author[c]{A.\,R.~Alhebsi}
\author[d]{M.~Alshamsi}
\author[e]{S. Alves Garre}
\author[g,f]{A. Ambrosone}
\author[h]{F.~Ameli}
\author[i]{M.~Andre}
\author[j]{L.~Aphecetche}
\author[k]{M. Ardid}
% ...
\affiliation[a]{INFN, Sezione di Catania, (INFN-CT) Via Santa Sofia 64, Catania, 95123 Italy}
\affiliation[b]{Universit{\'e}~de~Strasbourg,~CNRS,~IPHC~UMR~7178,~F-67000~Strasbourg,~France}
\affiliation[c]{Khalifa University of Science and Technology, Department of Physics, PO Box 127788, Abu Dhabi,   United Arab Emirates}
\affiliation[d]{Aix~Marseille~Univ,~CNRS/IN2P3,~CPPM,~Marseille,~France}
\affiliation[e]{IFIC - Instituto de F{\'\i}sica Corpuscular (CSIC - Universitat de Val{\`e}ncia), c/Catedr{\'a}tico Jos{\'e} Beltr{\'a}n, 2, 46980 Paterna, Valencia, Spain}
\affiliation[f]{INFN, Sezione di Napoli, Complesso Universitario di Monte S. Angelo, Via Cintia ed. G, Napoli, 80126 Italy}
\affiliation[g]{Universit{\`a} di Napoli ``Federico II'', Dip. Scienze Fisiche ``E. Pancini'', Complesso Universitario di Monte S. Angelo, Via Cintia ed. G, Napoli, 80126 Italy}
\affiliation[h]{INFN, Sezione di Roma, Piazzale Aldo Moro 2, Roma, 00185 Italy}
\affiliation[i]{Universitat Polit{\`e}cnica de Catalunya, Laboratori d'Aplicacions Bioac{\'u}stiques, Centre Tecnol{\`o}gic de Vilanova i la Geltr{\'u}, Avda. Rambla Exposici{\'o}, s/n, Vilanova i la Geltr{\'u}, 08800 Spain}
\affiliation[j]{Subatech, IMT Atlantique, IN2P3-CNRS, Nantes Universit{\'e}, 4 rue Alfred Kastler - La Chantrerie, Nantes, BP 20722 44307 France}
\affiliation[k]{Universitat Polit{\`e}cnica de Val{\`e}ncia, Instituto de Investigaci{\'o}n para la Gesti{\'o}n Integrada de las Zonas Costeras, C/ Paranimf, 1, Gandia, 46730 Spain}
% ...
\affiliation[be]{Universit{\'e} de Haute Alsace, rue des Fr{\`e}res Lumi{\`e}re, 68093 Mulhouse Cedex, France}
% ...


\maketitle
\flushbottom

\section{Introduction}
\label{sec:intro}

The discovery of neutrino oscillations at the turn of the century, implying the existence of at least two non-zero neutrino mass states, constituted the first landmark observation of a deviation from the Standard Model (SM) of particle physics \cite{RevModPhys.88.030501}. While most of the parameters that govern the oscillation mechanism are now measured with good precision \cite{Esteban:2020cvm}, some of them remain uncertain, such as the Dirac CP phase ($\delta$), the octant of the mixing angle $\theta_{23}$, and the neutrino mass ordering, which is either normal (NO: $m_1 < m_2 \ll m_3$) or inverted (IO: $m_3 \ll m_1 < m_2$). Other signatures of physics beyond the Standard Model (BSM) may exist in the neutrino sector and are actively searched for, one of them being the possibility of neutrino decay. 

The concept of neutrino decay was introduced in 1972 as a potential solution for the solar neutrino problem \cite{Bahcall:1972}, although subsequent investigations revealed that the neutrino decay scenario cannot account for the observed deficit of solar neutrinos \cite{Acker_1994}. Massive neutrinos may have radiative decay channels, e.g. $\nu_j\rightarrow \nu_i + \gamma$~\cite{Petcov:1976ff, MARCIANO1977303}; however the expected lifetimes for such processes are too long to be tested experimentally~\cite{ PhysRevD.25.766, PhysRevD.28.1664}. A variety of BSM 
theoretical scenarios have also been proposed, typically allowing for the decay of (Majorana or Dirac) neutrinos into a lighter fermion state and some BSM particles; we refer the reader to ref.~\cite{Arguelles:2022tki} for a brief overview of such, and more exotic, models.  

Neutrino decays can be phenomenologically classified into visible or invisible channels, depending on the nature of the decay products and their detectability \cite{Lindner_2001}. Invisible decay occurs when the decay products remain undetected, either because they are sterile neutrinos or because they have such low energy that they fall below the detection threshold of the experiment. In contrast, visible decay involves either the emission of photons or the production of lower-energy active neutrinos, which retain enough energy to be detected through interactions. In this paper, we primarily focus on the scenario of invisible neutrino decay, regardless of the specific decay model or products. This is mainly motivated by the fact that the expected decay signal from the non-sterile channels in the atmospheric neutrino telescopes is expected to contribute only negligibly to the event rate at the lowest detectable energy threshold because of a sharply-decreasing atmospheric neutrino flux with energy.

The invisible decay of relativistic neutrinos can be described by a depletion factor $D = e^{-\frac{m_i L}{\tau_i E}}$, where $\tau_i$ is the rest-frame lifetime of the neutrino mass state $m_i$, representing the fraction of neutrinos with energy $E$ that survive after travelling a distance $L$. The invisible neutrino decay is then characterised by the parameter $\alpha_i = \frac{m_i}{\tau_i}$, which is expressed in natural units of $\mathrm{eV^2}$. In principle, any of the neutrino mass eigenstates $\nu_1, \nu_2$ and $\nu_3$ having non-zero mass could potentially lead to invisible decays. 

Currently, the most stringent, albeit model-dependent, limits on neutrino decay come from cosmology \cite{Barenboim:2020vrr}. Other studies have shown that the decays of $\nu_1$ and $\nu_2$ are strongly restricted by data from the supernova SN1987A \cite{ FRIEMAN1988115} and solar neutrino experiments \cite{Bandyopadhyay_2003}. 
While the decay of $\nu_3$ remains an open possibility, ongoing accelerator, atmospheric, and reactor neutrino experiments have not yet observed any evidence of this phenomenon. Existing constraints on the neutrino lifetime based upon invisible decays are generally weaker, mainly arising from combined fits using data from T2K and NO$\nu$A \cite{NovaT2K} ($\alpha_3 < 2.4\times 10^{-4}~\mathrm{eV^2}$ at 90\% CL),  T2K and MINOS \cite{MinosT2K} ($\alpha_3 < 2.9\times 10^{-4}~\mathrm{eV^2}$ at 90\% CL), T2K,  MINOS and NO$\nu$A \cite{MINOST2KNOVA} ($\alpha_3 < 2.7\times 10^{-5}~\mathrm{eV^2}$ at 90\% CL) and SK, K2K and MINOS  \cite{SKK2KMINOS} ($\alpha_3 < 2.3\times 10^{-6}~\mathrm{eV^2}$ at 90\% CL).  Constraints on neutrino decay can also be inferred from astrophysical neutrino data. In particular,
the tension observed between different classes of events in IceCube data has been reduced by  $\sim 3\,\sigma$  when considering the invisible neutrino decay hypothesis~\cite{Denton:2018aml}.  For a more comprehensive exploration of the current status and future prospects concerning neutrino decay, both invisible and visible, readers can refer to the review in ref. \cite{Arguelles:2022tki}.

Due to their sensitivity in a large range of neutrino energies and propagation distances, atmospheric neutrino experiments are particularly appealing to investigate the effects of neutrino decay. Large-volume Cherenkov detectors, such as the KM3NeT/ORCA underwater neutrino telescope \cite{Adri_n_Mart_nez_2016}, offer the possibility to detect atmospheric neutrinos in the GeV to TeV range, which are an excellent probe for this kind of BSM effects that depend on the $L/E$ ratio. The potential of the complete ORCA detector for probing the invisible decay model using such atmospheric neutrinos has been studied in ref.~\cite{KM3NeT:2023ncz}, showing that it could improve  current limits on $\alpha_3$ by up to two orders of magnitude. 

In this paper, the invisible neutrino decay of $\nu_3$ is probed with the data sample collected with ORCA6, an early subarray of the KM3NeT/ORCA detector, in the period from January 2020 to November 2021. 
Section~\ref{sec:oscframe} briefly discusses the effects of invisible decay on neutrino oscillations. Section~\ref{sec:detector} presents the KM3NeT/ORCA neutrino detector. Section~\ref{sec:sample} summarises the event selection, reconstruction and classification procedures. The analysis method is described in section~\ref{sec:analysis}, and results are presented and discussed in section~\ref{sec:results}.

\section{Invisible decay effects on neutrino oscillations}
\label{sec:oscframe}

In the standard, 3-neutrino oscillation framework, flavour eigenstates   $\nu_{\beta}$  $(\beta=e, \mu, \tau$) are linearly related to the mass eigenstates $\nu_i$ $(i=1, 2, 3)$: $\nu_{\beta}=\sum_{i=1}^3 U^*_{\beta i} \nu_i$, where the $U_{\beta i}$ are the elements of the Pontecorvo-Maki-Nakagawa-Sakata (PMNS) mixing matrix~\cite{Pontecorvo1957MesoniumAA, Maki:1962mu}, which can be parametrised in terms of three real mixing angles $\theta_{ij}\ (i<j)$  and one complex phase $\delta$ accounting for possible CP-violating effects. In the case of atmospheric neutrinos crossing the Earth, coherent scattering on electrons along the path of the neutrino modifies the oscillation probabilities. The main effect is an amplification of the transition probabilities ($\nu_{\mu} \leftrightarrow \nu_{e}$) in the resonance region (around $3 - 8$ GeV) \cite{Smirnov:2003da}.

The Hamiltonian that describes the propagation of a neutrino through matter, incorporating the effect of invisible decay, can be written as:

\begin{equation}
H_{\text{Total}}=\frac{1}{2E} \left( H_{0}+H_{D}+H_M \right),
\end{equation}

\noindent where $E$ is the neutrino energy, $H_0$ represents the Hamiltonian in vacuum, $H_M$ encompasses the effects of coherent scattering on electrons along the path of the neutrino, and $H_D$ accounts for neutrino decay. In the flavour eigenstate basis, the Hamiltonian responsible for oscillations reads:

\begin{equation}
    \centering
   H_{\text{Total}}=\frac{1}{2E} \left[U
  \left( {\begin{array}{ccc}
   0 & 0 & 0 \\
   0  & \Delta m^2_{21} & 0 \\
   0 & 0 & \Delta m^2_{31} \\
  \end{array} } \right)U^{\dagger}+ U
  \left( {\begin{array}{ccc}
   0 & 0 & 0 \\
   0  & 0 & 0 \\
   0 & 0 & -i \alpha_3 \\
  \end{array} } \right)U^{\dagger}  \right]+  
  \left( {\begin{array}{ccc}
   V & 0 & 0 \\
   0  & 0 & 0 \\
   0 & 0 & 0 \\
  \end{array} } \right),
\end{equation}

\noindent with $U$ being the PMNS matrix, $V=\pm\sqrt{2}N_eG_F$ the matter potential, $N_e$ the electron density in matter and $G_F$ the Fermi constant.

The introduction of $H_D$ thus essentially leads to a shift in the mass  splitting, from $\Delta m_{31}^2$ to $\Delta m_{31}^2-i \alpha_3$, becoming a complex number. Neutrino decay induces a global reduction in the neutrino flavour oscillation probabilities and introduces a damping effect that reduces the amplitude of the oscillatory terms \cite{Chattopadhyay:2021eba,Chattopadhyay:2022ftv}. 

\begin{figure}[H]
 \centering
  \subfloat{
   \label{Fig2_Osc-1}
     \includegraphics[width=0.48\textwidth]{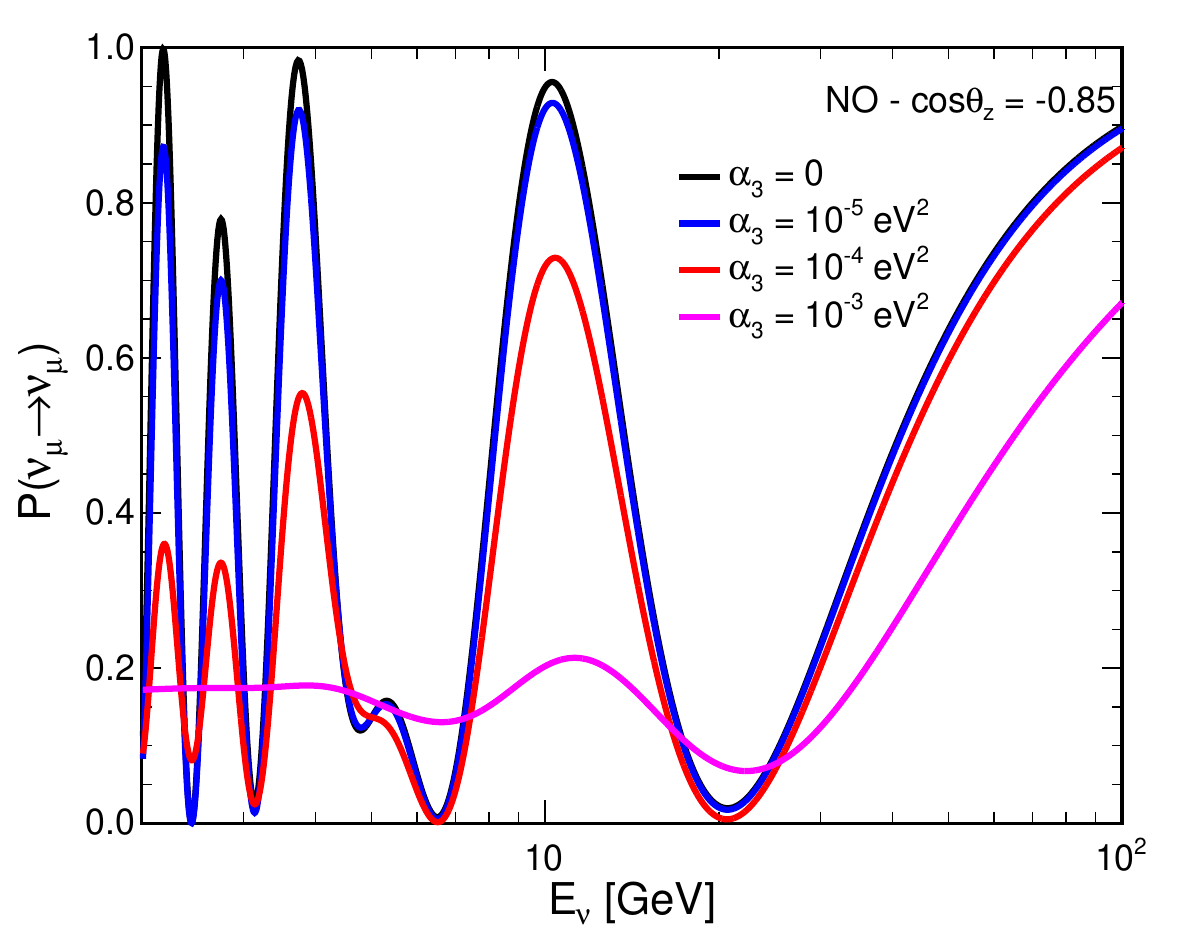}}
  \subfloat{
   \label{Fig2_Osc-2}
      \includegraphics[width=0.48\textwidth]{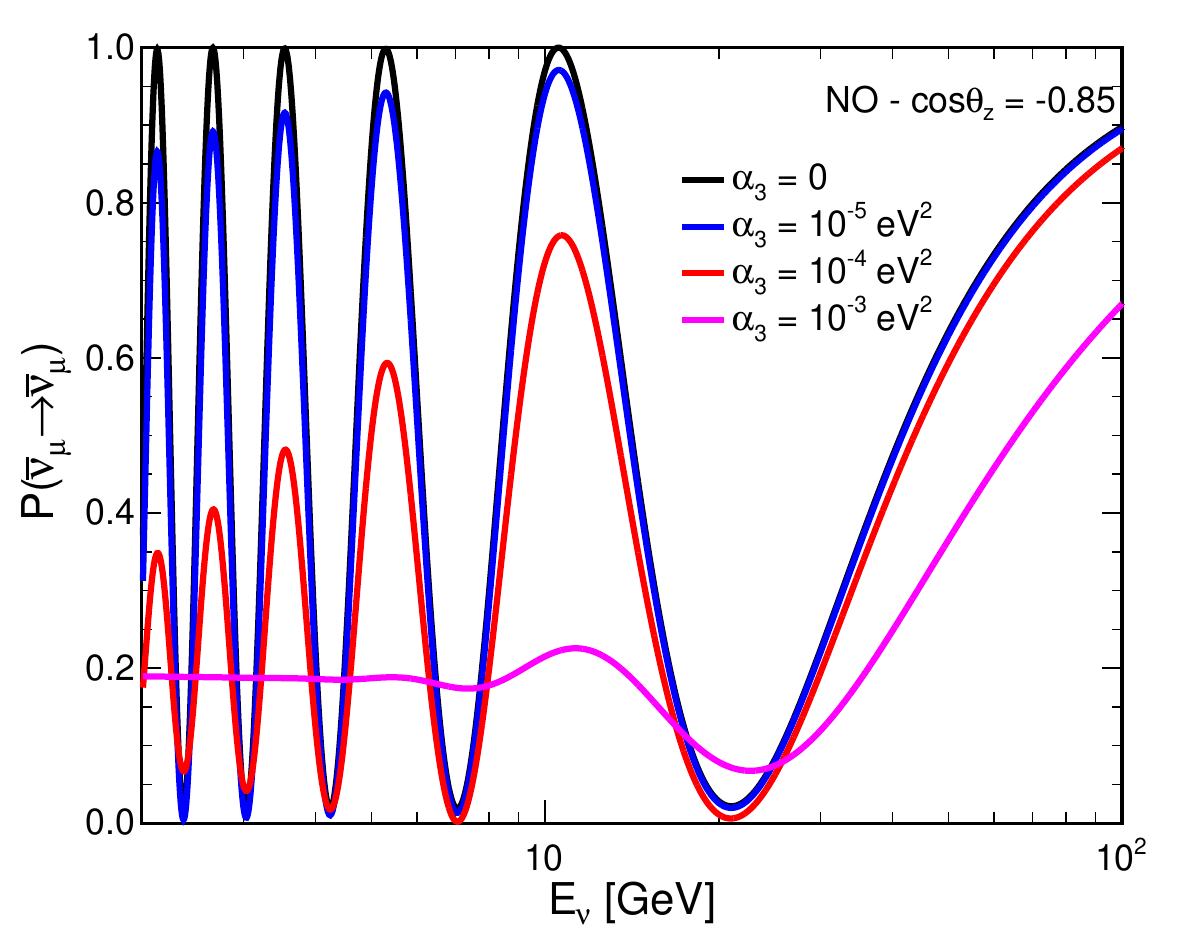}} \\
       \subfloat{
   \label{Fig2_Osc-3}
     \includegraphics[width=0.48\textwidth]{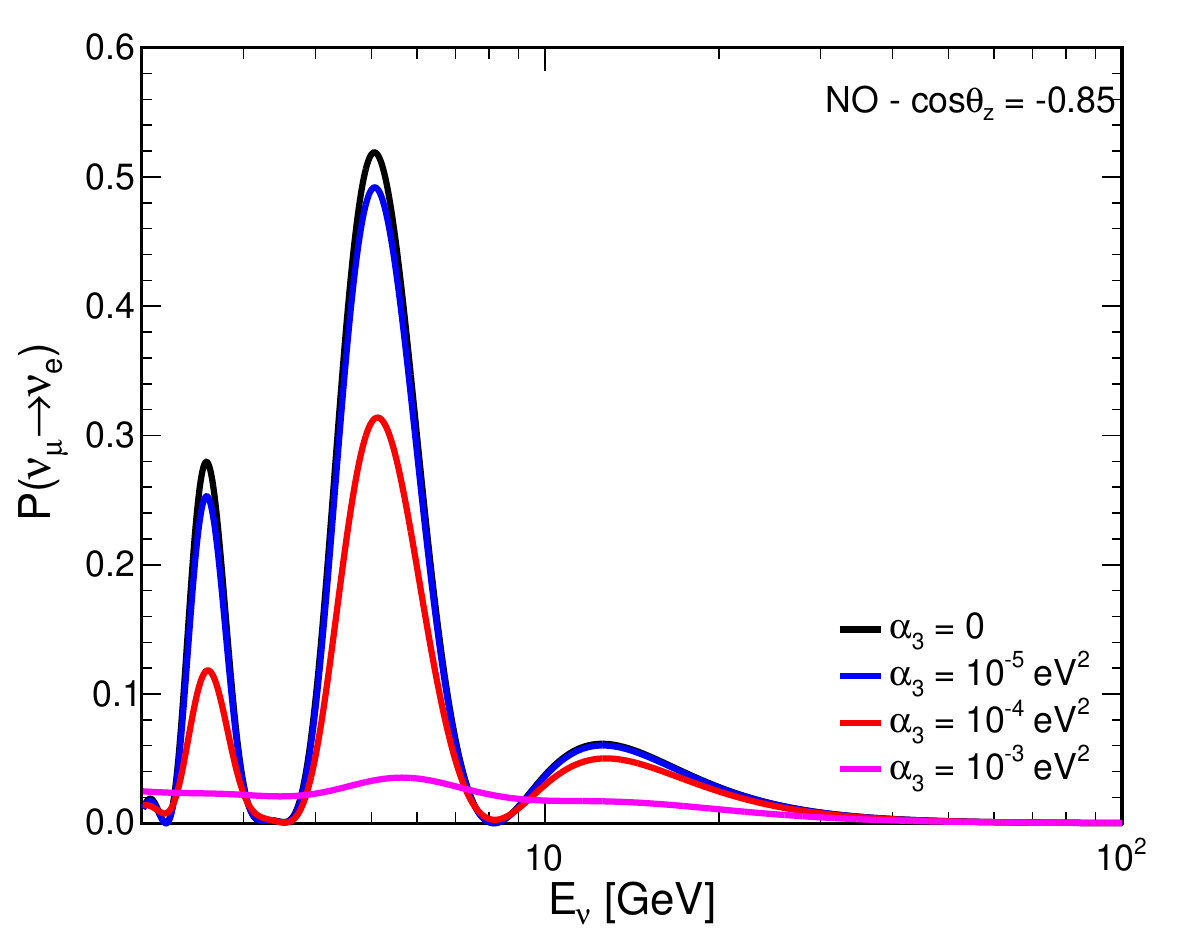}}
  \subfloat{
   \label{Fig2_Osc-4}
      \includegraphics[width=0.48\textwidth]{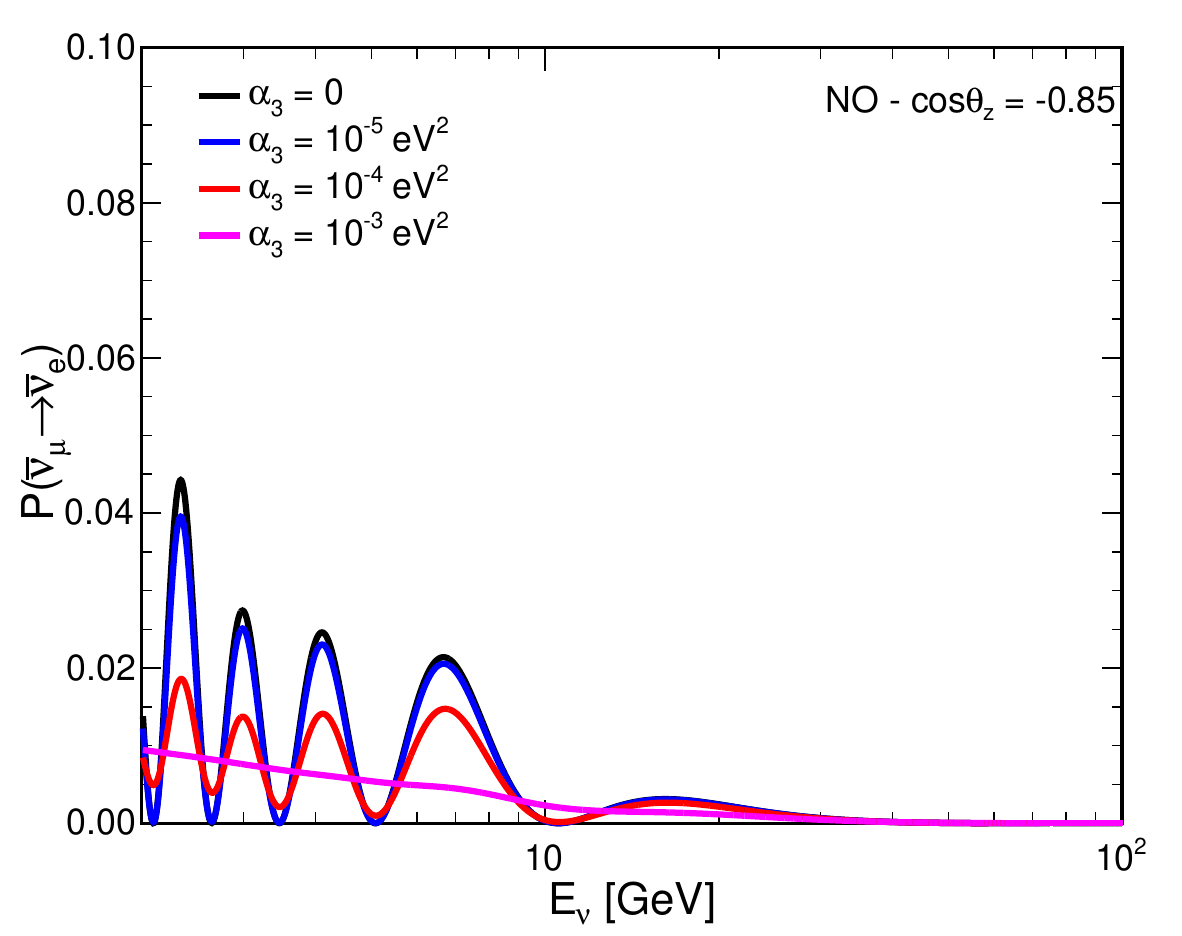}}\\
       \subfloat{
   \label{Fig2_Osc-3}
     \includegraphics[width=0.48\textwidth]{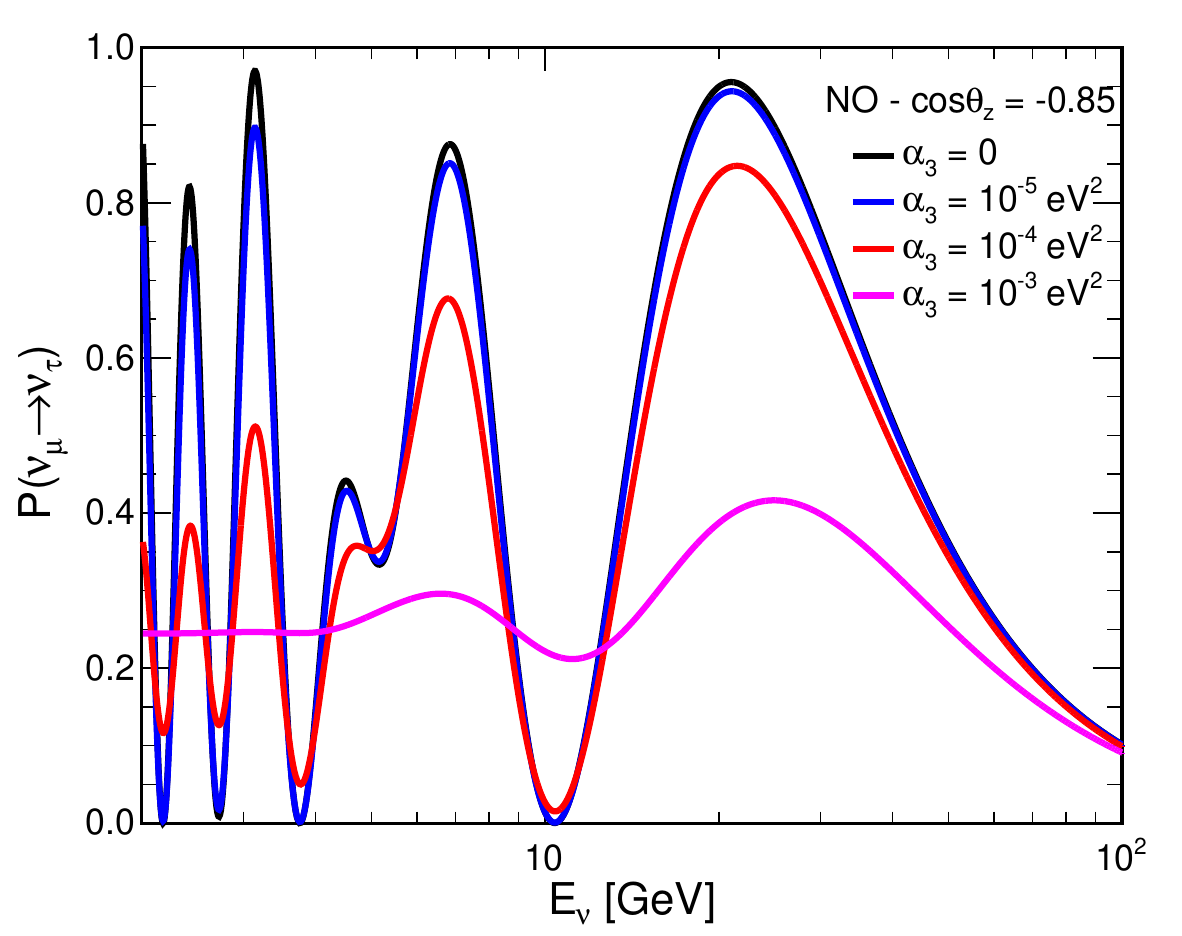}}
  \subfloat{
   \label{Fig2_Osc-4}
      \includegraphics[width=0.48\textwidth]{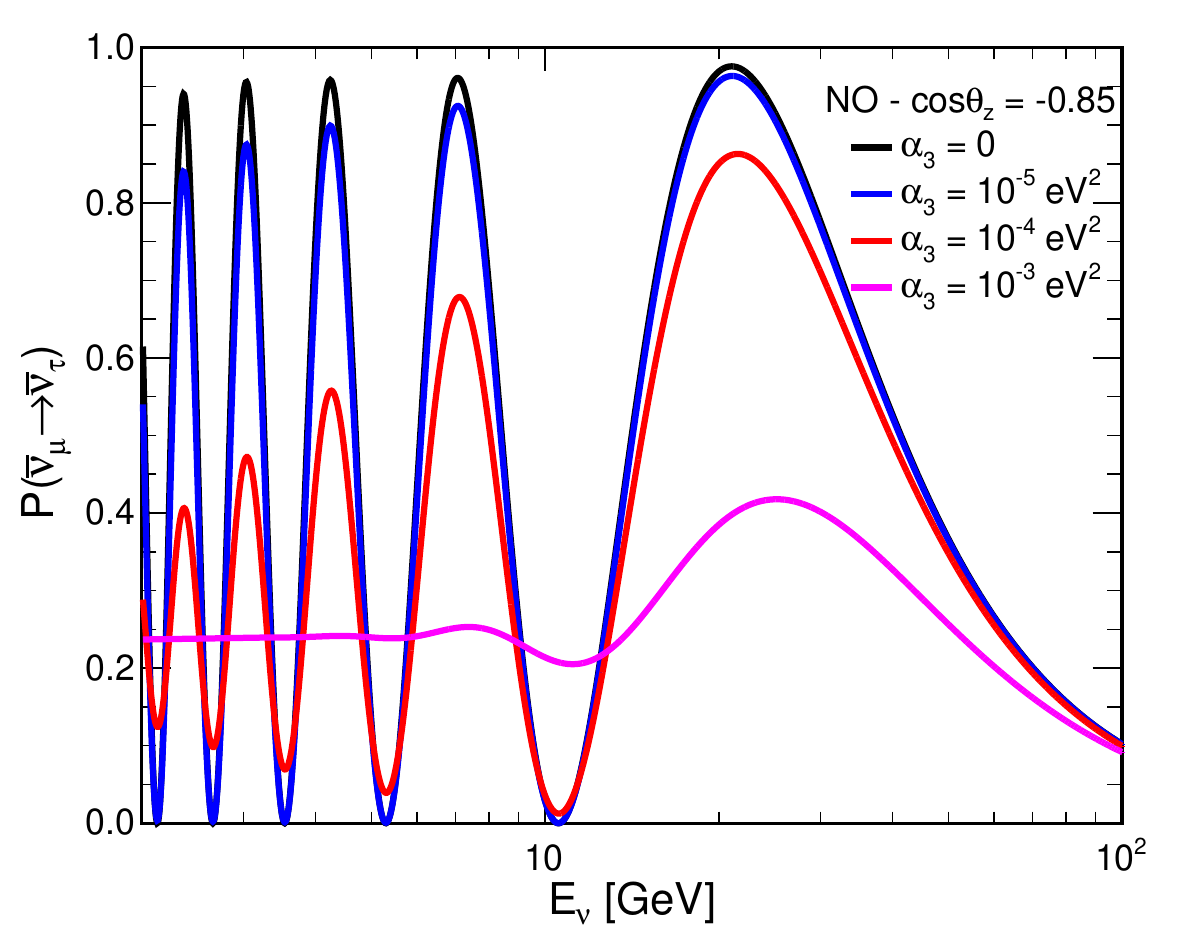}}

 \caption{Probability of muon neutrino survival (top left), muon-to-electron neutrino transition (middle left), muon-to-tau neutrino transition (bottom left), muon antineutrino survival (top right), muon-to-electron antineutrino transition (middle right) and muon-to-tau antineutrino transition (bottom right) as a function of energy. All plots are obtained assuming normal ordering (NO) and $\cos \theta_z = - 0.85$, where $\theta_z$ is the zenith angle associated with the neutrino trajectory, as measured from the vertical at detector location. Four values of the decay parameter are  considered:  $\alpha_3 = 0$ (black),  $\alpha_3 = 10^{-5}~\mathrm{eV^2}$ (blue), $\alpha_3 =  10^{-4}~\mathrm{eV^2}$ (red) and $\alpha_3 =  10^{-3}~\mathrm{eV^2}$ (magenta). Oscillation parameters are set to the NuFit 5.0 values with SK data \cite{Esteban:2020cvm}. }
 \label{Osc3}
\end{figure}

\begin{figure}[H]
 \centering
  \subfloat{
   \label{Fig3_Cor1}
     \includegraphics[width=1.0\textwidth]{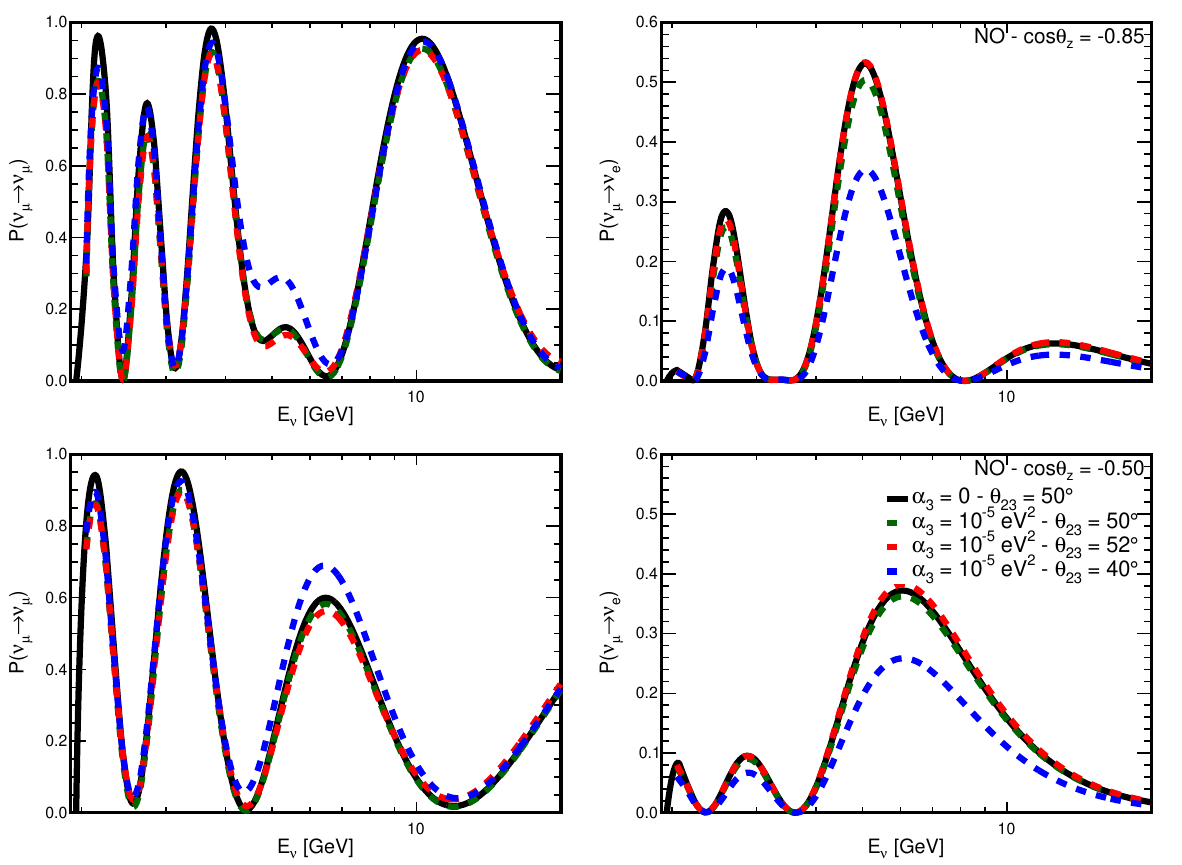}}

 \caption{Muon neutrino survival (left) and electron-to-muon transition (right) probabilities as a function of energy, at a cosine of the zenith angle $\cos \theta_z = - 0.85$ (top) and $\cos \theta_z = - 0.50$ (bottom), assuming NO. Four cases are shown: $\alpha_3 = 0$ with $\theta_{23}=50^{\circ}$ (solid black), and $\alpha_3=10^{-5}~\mathrm{eV^2}$ 
 with $\theta_{23} =50^\circ$ (dashed green),  $\theta_{23}=52^{\circ}$ (dashed red), and $\theta_{23}=40^{\circ}$ (dashed blue). 
 }
 \label{Th23Change}
 \end{figure}

The influence of neutrino decay on the oscillation patterns of Earth-crossing neutrinos is illustrated in figure~\ref{Osc3} 
for a specific neutrino path length and different values of the $\alpha_3$ parameter. Neutrino oscillation probabilities are computed with the OscProb software \cite{joao_coelho_2023_10104847} and the Preliminary Reference Earth Model \cite{PREM} is used for the density profile of the Earth's interior. The effects of decay due to $\alpha_3$ are more pronounced in channels related to the muon flavour, primarily because $\nu_3$ contributes more to $\nu_{\mu}$ compared to $\nu_e$. In vacuum, the neutrino decay affects most significantly the $P_{\mu\mu}$ channel, regardless of the mass ordering. However, the amplification of the transition probabilities
$\nu_{\mu} \leftrightarrow \nu_{e}$ (resp. $\bar{\nu}_{\mu} \leftrightarrow \bar{\nu}_{e}$) for the normal ordering (resp. inverted ordering) 
by matter effects renders these channels more sensitive to neutrino decay than in the vacuum case, especially in the energy range around the resonance. 
Neutrino decay also affects $\nu_{\mu} \leftrightarrow \nu_{\tau}$  transitions, with a more pronounced effect at energies below 10 GeV. Given the kinematic threshold for tau production, this channel plays a less significant role in overall sensitivity to neutrino decay.\\
\indent The parameters $\alpha_3$ and $\theta_{23}$ exhibit subtle correlation effects that vary between oscillation and survival channels. 
The interplay between these two parameters has been extensively investigated for specific baselines in previous studies, such as refs. \cite{ESSnuSB_choubey2020exploring, Dey_2024, FirstSecondMin, MINOST2KNOVA}. Decay-induced attenuation in transition channels can be mimicked by a standard oscillation scenario with a lower $\theta_{23}$ value, while in survival channels, the effect is reversed. Figure \ref{Th23Change} shows the muon neutrino survival and electron-to-muon transition probability for four 
combinations of $\theta_{23}-\alpha_3$ values and two different incoming directions. The decrease of the survival probability $P_{\mu\mu}$ at the oscillation maxima due to decay effects could be partially compensated by reducing the value of $\theta_{23}$ to the lower octant, but this would increase the probability in the energy range where matter effects are relevant. 
On the other hand, in the case of the transition probability $P_{e\mu}$, a higher value of $\theta_{23}$ 
can compensate the decrease caused by the decay effects. The change in the value of $\theta_{23}$ needed to compensate these effects depend on the cosine of the zenith angle, since the decay effects would be larger the longer the path. Therefore, this interplay degrades sensitivity to the invisible decay parameter unless the experiment can resolve a wide range of baselines and oscillation channels, allowing these effects to be disentangled \cite{KM3NeT:2023ncz}.

\section{The KM3NeT/ORCA detector}
\label{sec:detector}

The KM3NeT research infrastructure comprises two water Cherenkov detectors, called KM3NeT/ARCA and  KM3NeT/ORCA, both currently under construction in the Mediterranean Sea \cite{Adri_n_Mart_nez_2016}. KM3NeT/ARCA (Astroparticle Research with Cosmics in the Abyss) is located off the Sicilian coast near Capo Passero (Italy), 100~km offshore, at a depth of 3500~m. Its design is optimised for the detection of high-energy neutrinos from astrophysical sources in the TeV$-$PeV range. KM3NeT/ORCA (Oscillation Research with Cosmics in the Abyss) is being built near Toulon (France), 40~km offshore, at a depth of 2500~m, with the objective of measuring atmospheric neutrino oscillations in the GeV$-$TeV range and determining the neutrino mass ordering \cite{NMOpaper}. 

Both ARCA and ORCA detectors consist of arrays of detection units (DUs), which are vertical lines anchored to the seabed and supporting 18 digital optical modules (DOMs),  pressure-resistant glass spheres, each equipped with 31 photomultiplier tubes (PMTs) along with the associated readout electronics and calibration instruments~\cite{KM3NeT:2022pnv}.  Upon completion, KM3NeT/ORCA will comprise 115 detection units, with an average horizontal spacing between DUs of $\sim20$~m and a vertical inter-DOM spacing of  $\sim9$~m, corresponding to a total instrumented volume of about  $7\times 10^6~\mathrm{m^3}$. KM3NeT/ORCA has already begun data collection during the construction phase, using increasing subsets of the planned final detector setup. During the initial phase referred to as ORCA6, the detector was operating with six out of the total 115 DUs, corresponding to an instrumented volume of about $4\times 10^5~\mathrm{m^3}$.

The detection principle is based on the observation of Cherenkov radiation induced
by relativistic charged particles originating from neutrino interactions and travelling in seawater. When PMT pulses exceed the predefined threshold, they are digitised and analysed based on their start time and duration. The data are used to reconstruct the energy and direction of the incoming neutrino. More details about the KM3NeT data acquisition and event reconstruction are provided in refs. \cite{Adri_n_Mart_nez_2016, km3netcollaboration2024measurementneutrinooscillationparameters}.

\section{Data sample and event selection}
\label{sec:sample}
This work uses data collected with the partial configuration ORCA6 of the KM3NeT/ORCA detector. In this section, the event selection and classification 
procedures are briefly summarised; a complete description can be found in ref.~\cite{km3netcollaboration2024measurementneutrinooscillationparameters}
 since the same dataset is used.

The data analysed were collected between January 2020 and November 2021 using only period characterised by stable environmental conditions. After this selection, a total of 510 days out of the 633 days of data taking remain, corresponding to a total 
exposure of 433 kton-years\footnote{The exposure is computed by summing, for all the selected data taking periods, the product of the livetime by the number of active PMTs, and assuming that each active PMT instruments 108.8 tons of water (obtained by dividing the planned instrumented mass of KM3NeT/ORCA115 by the total number of PMTs in the full detector).}. The event reconstruction considers two distinct topologies: track-like and shower-like. Track-like events arise from charged-current (CC) interactions of  
$\nu_{\mu}$  and 
$\nu_{\tau}$ which produce a muon in the final state. Events with electromagnetic or hadronic showers are produced by
$\nu_e$ and 
$\nu_{\tau}$ CC interactions, as well as neutral-current (NC) interactions of all neutrino flavours.  Quality cuts based on the reconstruction filter out noise events, primarily from $^{40}$K decays and bioluminescence \cite{Albert_2018, annurev-marine-120308-081028}. The analysis focuses on up-going events with the cosine of the zenith angle $\cos\theta_z<0$ to minimise atmospheric muon background. To further distinguish neutrino-induced signals from misidentified atmospheric muons, a boosted decision tree (BDT) trained on reconstruction features is applied. The final event selection includes 5828 events, with an atmospheric muon contamination below 2\%.

A second BDT is employed to classify events into track-like and shower-like categories based on their topologies. The track-like category is then divided into two subcategories: (i) a high-purity track-like class with minimal atmospheric muon contamination and an estimated  $\nu_{\mu}$ CC purity of 95\% and (ii) a low-purity track-like class with less than 4\% muon contamination and 90\% $\nu_{\mu}$ CC purity. The classification thresholds are optimised to maximise sensitivity to standard neutrino oscillations. In table~\ref{Table:ORCA6Selection_w}, the observed and expected number of selected events per class and interaction channel are shown. 

This division of track-like events into high- and low-purity classes enhances sensitivity to standard oscillation parameters by isolating events with superior angular resolution in the high-purity class. The energy range for events in the track classes spans from 2 to 100 GeV, while the shower class covers a broader range from 2 GeV to 1 TeV. This energy threshold for track-like events is applied to minimise the number of high-energetic tracks that pass through the detector and just leave a small fraction of their energy in it. The track energy is essentially estimated from the measured track length. Due to the limited size of the detector the algorithm provides a constantly lower energy measurement for neutrino energies above 50 GeV as most tracks at these energies deposit only a fraction of their total energy within the detector.

\clearpage
\begin{table}[H]
KM3NeT/ORCA6, 433 kton-years \\
\begin{tabular}{|c|rrr|r|}

\hline
    Channel         & \textbf{High Purity Tracks} & \textbf{Low Purity Tracks} & \textbf{Showers} & \textbf{Total} \\ \hline \hline
$\nu_\mu$ CC                   & 1166.2               & 1202.9               & 639.6              & 3008.7           \\
$\bar{\nu}_\mu$ CC             & 607.3                & 605.6                & 217.8              & 1430.7           \\
$\nu_e$ CC                     & 40.2                & 69.4                  & 457.7              & 567.3            \\
$\bar{\nu}_e$ CC               & 16.7                 & 26.9                 & 190.6              & 234.2            \\
$\nu_\tau$ CC                  & 14.6                 & 13.8                  & 95.8               & 124.2            \\
$\bar{\nu}_\tau$ CC            & 6.6                  & 6.0                  & 37.2               & 49.8             \\
$\nu$ NC                       & 10.4                 & 18.5                 & 236.3              & 265.2            \\
$\bar{\nu}$ NC                 & 3.2                  & 5.5                  & 70.2               & 78.9             \\
Atm. Muons                     & 4.3                  & 53.3                 & 13.5               & 71.1           \\ \hline
\textbf{Total MC}              & 1869.5               & 2001.9               & 1958.7             & 5830.1           \\ \hline \hline
\textbf{Total Data}            & 1868               & 2002               & 1958             & 5828           \\ \hline
\end{tabular}
\caption{Number of  expected events and data events passing the selection criteria per class and interaction channel compared to the total number of data events in each class for an  exposure of 433 kton-year. The values correspond to the best fit which is detailed in section~\ref{sec:analysis} with nuisance parameters as given in table~\ref{MINOS_Data_DD}.}
\label{Table:ORCA6Selection_w}
\end{table}

\section{Analysis method}
\label{sec:analysis}
The analysis procedure to constrain the insivible decay parameter $\alpha_3$ is based on the minimisation of a binned negative log-likelihood of the 2-dimensional distribution of events in reconstructed energy $E_{\text{reco}}$ and cosine of the zenith angle $\cos\theta_{\text{reco}}$ comparing the observed data to a model prediction~\cite{Cousins2013GeneralizationOC, BAKER1984437}:
\begin{align}
\lambda(\alpha_3;\vec{\epsilon}) = -2\ln L= & 2 \sum_{i}  \Biggl[ (\beta_{i} N^{\text{mod}}_{i}(\alpha_3 ; \vec{\epsilon})-N^{\text{dat}}_{i})+N^{\text{dat}}_{i} \ln \left( \frac{N^{\text{dat}}_{i}}{\beta_{i}N^{\text{mod}}_{i}(\alpha_3;\vec{\epsilon})}\right)\Biggr] \nonumber\\& \qquad +\frac{(\beta_{i}-1)^2}{\sigma_{\beta i}^2}  +  \sum_k \left(\frac{\epsilon_k-\langle\epsilon_k\rangle}{\sigma_{\epsilon k}}\right)^2, 
    \label{ORCA6_eq_sat}
\end{align}
\noindent  where $N^{\text{mod}}_{i}$ and $N_{i}^{\text{dat}}$ are respectively the expected and observed number of events in bin $i$. The vector $\vec{\epsilon}$ represents the nuisance parameters, some of them externally constrained by other experiments. This information enters the log-likelihood as a Gaussian term derived from the probability density function (PDF) of the auxiliary measurement. The mean value $\langle \epsilon_k\rangle$ and standard deviation $\sigma_{\epsilon k}$ are the parameters used to define these PDFs. The coefficients 
$\beta_i$  are normally distributed with uncertainties $\sigma_i$, which account for the statistical uncertainties arising from limited Monte Carlo statistics, following the Barlow-Beeston light approach \cite{BARLOW1993219, Conway:1333496, km3netcollaboration2024measurementneutrinooscillationparameters}.

The number of expected events is computed using the KM3NeT package Swim \cite{Bourret:2018kug}, which combines the atmospheric neutrino flux (HONDA 2014 at the Frejus site without mountain over the detector for solar minima \cite{HondaFlux}), neutrino oscillation probabilities (OscProb package \cite{joao_coelho_2023_10104847}) and the detector response weighted by the respective neutrino-nucleon CC and NC cross sections (GENIE \cite{andreopoulos2015genie,Aiello_2020}). The detector response is simulated with Monte Carlo events by mapping the expected rate at a given bin of reconstructed energy and reconstructed cosine of the zenith angle to the corresponding bins of true energy and true cosine of the zenith angle.  Neutrino events in KM3NeT/ORCA are generated with gSeaGen \cite{Aiello_2020} and atmospheric muons are generated with MUPAGE \cite{Carminati:2008qb, Becherini:2005sr}. Cherenkov photons induced by charged particles are propagated to the PMTs using the KM3NeT package KM3Sim \cite{Tsirigotis:2011zza}. To speed up the simulation of light propagation in the case of high-energy particles, a custom KM3NeT package based on precomputed tables of PDFs of the light arrival time is used. The optical backgrounds due to the PMT dark count rate and to the decay of $^{40}\mathrm{K}$ present in seawater are included through a KM3NeT package  which also simulates the digitised output of PMT responses and the readout. The simulated events then follow the same triggering and reconstruction chain as described in section~\ref{sec:detector} for real data.

Those oscillation parameters for which KM3NeT/ORCA has no sensitivity are fixed to NuFit 5.0 including SK data values and shown in table~\ref{Table2-NufitO6}. The nuisance parameters considered in this analysis can be cast into 3 categories: \textit{normalisation factors}, \textit{flux shape systematics} and the \textit{absolute energy scale} of the detector. The \textit{normalisation factors} aim to account for uncertainties in the cross sections and event selection efficiency and are applied to scale respectively the overall amount of events, $f_{\text{all}}$, the High Purity Tracks $f_{\text{HPT}}$ and Shower $f_{\text{S}}$ events, the NC events, $f_{\text{NC}}$, the $\tau$ CC events, $f_{\tau CC}$, the atmospheric muon events, $f_{\mu}$ and the high-energy events simulated with a different propagation software $f_{\text{HE}}$. The $f_{\text{HE}}$ normalisation factor is introduced to take into account the different assumptions on light propagation made by the two light propagation software packages used. A scaling is therefore applied for NC events with true energy above 100~GeV and for CC events with true energy above 500~GeV. The \textit{flux shape systematics} aim at modelling uncertainties in the neutrino flux by altering the ratio of up-going to horizontally-going neutrinos, $\delta_{\theta}$, the spectral index $\delta_{\gamma}$, the ratio of $\nu_{\mu}$ to $\bar{\nu}_{\mu}$, $s_{\mu\overline{\mu}}$, the ratio of $\nu_{e}$ to $\bar{\nu}_{e}$, $s_{e\overline{e}}$, and the ratio of $\nu_{e}$ to $\nu_{\mu}$, $s_{e\mu}$. The \textit{absolute energy scale} of the detector $E_{s}$ is implemented to account for uncertainties in the water properties (light absorption and scattering) and in PMT efficiencies, by shifting the true energy response function. Central values and uncertainties are summarised in table~\ref{MINOS_Data_DD}. The normalisations of the overall amount of events $f_{\text{all}}$, the High Purity Track $f_{\text{HPT}}$ and Shower $f_{\text{S}}$ events, and the atmospheric muon events $f_{\mu}$ are fitted freely without any constraint to accommodate for uncertainties in the selection efficiency. More details on the implementation of nuisance parameters are provided in ref. \cite{km3netcollaboration2024measurementneutrinooscillationparameters}.

\begin{table}[H]
\centering
    \begin{tabular}{|l|r|}
      \hline
      \textbf{Parameter} & \textbf{NO} \\
      \hline
      $\theta_{12}$ & $33.44^{\circ}$ \\
      \hline
      $\theta_{13}$  & $8.57^{\circ}$  \\
      \hline
      $\Delta m_{21}^2$ & $7.42 \times 10^{-5}~\mathrm{eV^2} $\\ 
      \hline
      $\delta_{\text{CP}}$ & $197^{\circ}$ \\
      \hline
     
    \end{tabular}
    \caption{The three-flavour neutrino oscillation parameters which are fixed in the analysis from NuFit 5.0 \cite{Esteban:2020cvm} for normal ordering including SK data.}
    \label{Table2-NufitO6}
    \end{table}
    

\section{Results}
\label{sec:results}

The best fit is obtained by minimising the negative log-likelihood ratio (equation  \ref{ORCA6_eq_sat}) using 8 starting points, $\theta_{23} =\{40^{\circ}, 50^{\circ}\}$, $\Delta m^2_{31} = \{ 2.517\times10^{-3}, -2.428\times10^{-3}\}~\mathrm{eV^2}$ and $E_s=\{0.95, 1.05\}$. In each fit, the parameter space is restricted to the corresponding $\theta_{23}$ octant, mass ordering and energy scale below/above one to avoid local minima.

\begin{figure}[]
    \centering    \includegraphics[width=0.75\linewidth]{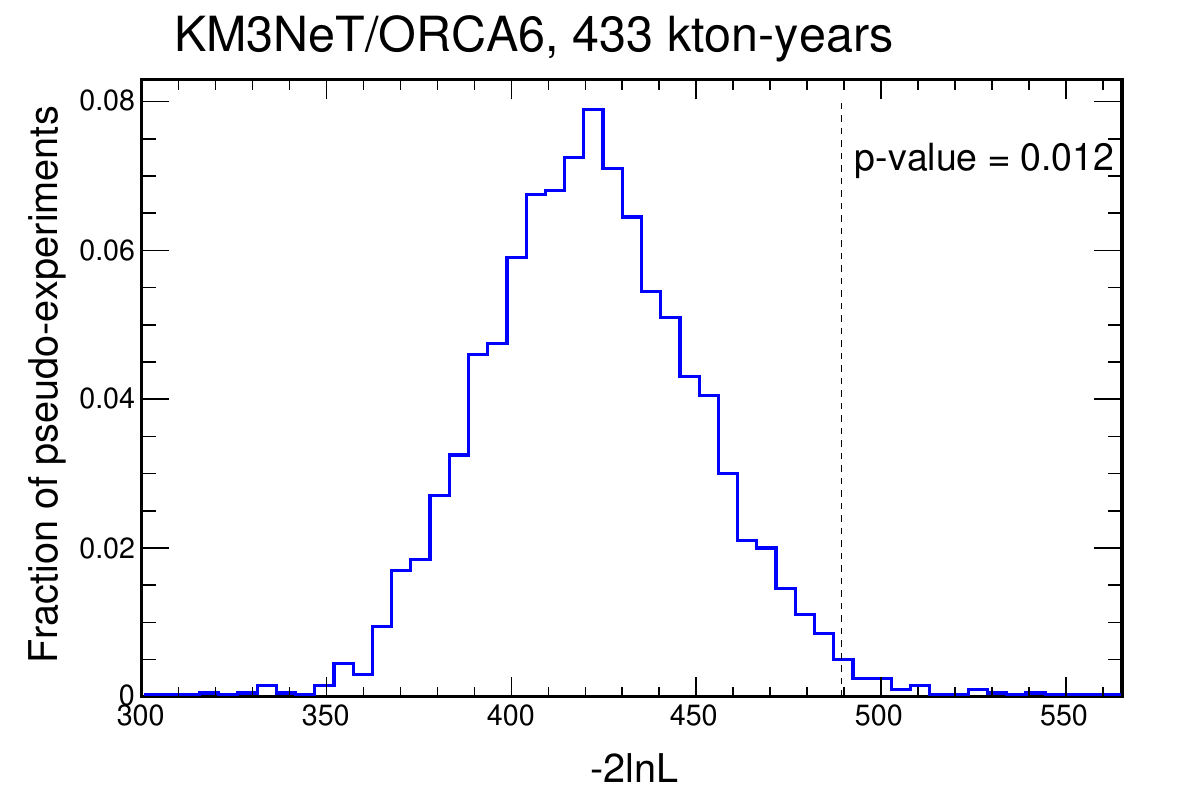}
    \caption{Distribution of $\lambda$ values from 2000 pseudo-experiments used to carry out the goodness-of-fit test. The vertical line indicates the observed value of $\lambda$ corresponding to the best fit to the data. }
    \label{Fig:C6-GoF}
\end{figure}

 After performing the full minimisation of equation \eqref{ORCA6_eq_sat}  the observed log-likelihood ratio $\lambda$ is found to be $489.3$. To assess the goodness-of-fit, a set of 2000 pseudo-experiments are generated assuming NuFit 5.0 values, $\alpha_3=0$ and the nuisance parameters at their nominal values. The log-likelihood ratio $\lambda_{\text{GoF}} = -2 \ln L$ is computed for each of them and the corresponding distribution is shown in figure~\ref{Fig:C6-GoF}. The probability of observing a $\lambda$ equal or larger than the observed value is $(1.21 \pm 0.24)\%$. The uncertainty is derived by bootstrapping the 2000 pseudo-experiments, i.e. sampling the pseudo-experiments with replacement to recompute the p-value. 

The best-fit value is $\alpha_3= 0.92^{+1.08}_{-0.57}\times 10^{-4}~ \mathrm{eV^2}$, corresponding to a scenario with $\theta_{23}$ in the second octant and NO. The errors are computed via the Feldman-Cousins (FC) method \cite{Feldman_1998} for 68\% CL. The best-fit values of the nuisance parameters, along with their post-fit uncertainties at 68\% CL are shown in table \ref{MINOS_Data_DD} assuming Wilks' theorem \cite{s__s__wilks_1938}. The post-fit uncertainties are computed by profiling each of them as a parameter of interest.

  \begin{table}[H]
 \centering
\begin{tabular}{|c|c|r|r|r|}
 \hline
  Parameter & CV $\pm$ uncert.&  Std. Best Fit & Decay Best Fit & Post-fit uncert.  \\ \hline
   $f_{\text{all}}$ &1.00 & 1.11 &1.35  & $-$0.20/+0.25 \\ \hline
   $f_{HPT}$ & 1.00 &0.92 & 0.92  & $-$0.04/+0.04 \\ \hline
   $f_{S}$ & 1.00 & 0.92 &0.88  & $-$0.06/+0.06 \\ \hline
   $f_{HE}$ & 1.0 $\pm$ 0.5 & 1.59 & 1.81  & $-$0.32/+0.35 \\ \hline
   $f_{\mu}$ &  1.00 & 0.51 &0.25  & $-$0.25/+0.35 \\ \hline
   $f_{\tau CC}$ &1.00 $\pm$ 0.20 & 0.92 &0.96  & $-$0.19/+0.19 \\ \hline
   $f_{NC}$ & 1.00 $\pm$ 0.20 & 0.86 &0.89  & $-$0.19/+0.19 \\ \hline
   $s_{\mu \bar{\mu}}$ & 0.00 $\pm$ 0.05& 0.00  & 0.00  & $-$0.05/+0.05 \\ \hline
   $s_{e \bar{e}}$ & 0.00 $\pm$ 0.07&0.01 & 0.01  & $-$0.07/+0.07 \\ \hline
   $s_{\mu e}$ & 0.00 $\pm$ 0.02 &-0.004& -0.004  & $-$0.020/+0.020 \\ \hline
$\delta_{\gamma}$ & 0.0 $\pm$ 0.3 & -0.019& -0.075  & $-$0.035/+0.04 \\ \hline
   $\delta_{\theta}$ & 0.000 $\pm$ 0.020 &-0.005& 0.004  & $-$0.020/+0.020 \\ \hline
   $E_s$ & 1.00 $\pm$ 0.09 & 1.03 &0.98  & $-$0.08/+0.10\\ \hline
    $\theta_{23}$ [$^\circ$]& 49.2 & 45.5 &46.2  & $-$4.0/+4.0 \\ \hline
   $\Delta m^{2}_{31} [\times 10^{-3}~\mathrm{eV^2}]$ & 2.52 & 2.18 &2.15  & $-$0.25/+0.28 \\ \hline
\end{tabular}
    \caption{
Nuisance and oscillation parameters considered in this analysis. The second column provides the central value and uncertainty of the Gaussian prior entering the log-likelihood ratio calculation. The third, fourth and fifth columns correspond respectively to the best-fit value assuming standard oscillation hypothesis, the best-fit value assuming decay scenario and the corresponding post-fit uncertainty at 68\% CL as obtained from the fit of ORCA6 data to $\alpha_3$. }
    \label{MINOS_Data_DD}
   \end{table}

\begin{figure}[]
 \centering
     \includegraphics[width=0.75\linewidth]{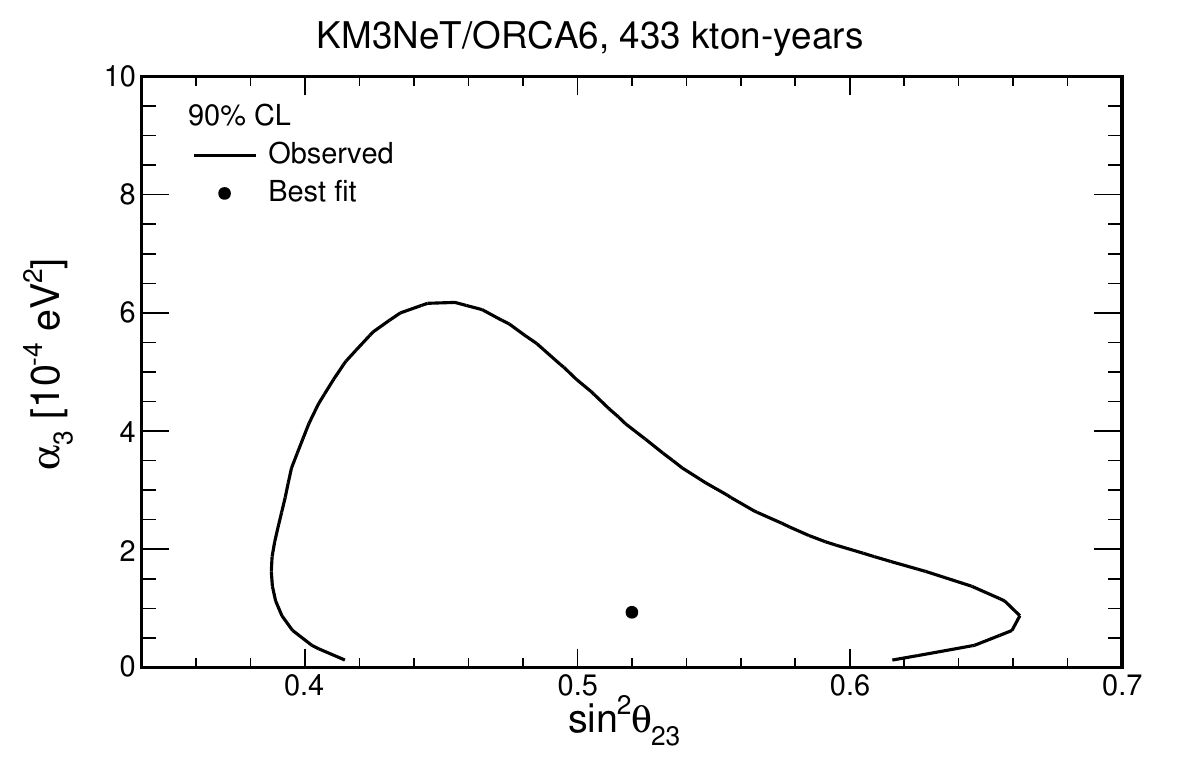}
    \caption{Allowed region at 90\% CL obtained from ORCA6 data for the $\theta_{23}-\alpha_3$ parameters.  The best-fit value is indicated with a dot. }
    \label{Fig:DataContour}
\end{figure}

The allowed region at $90\%$~CL  for both the mixing angle   $\theta_{23}$ and the invisible decay parameter $\alpha_3$ is shown in figure \ref{Fig:DataContour}, with the best-fit value indicated with a dot. The contour is derived assuming Wilks' theorem due to computing constraints. The shape illustrates the interplay between both parameters discussed in section~\ref{sec:oscframe}: the allowed range for $\alpha_3$ is larger in the first octant region. While the current sensitivity of ORCA6 is insufficient to disentangle the $\theta_{23}$ octant,  this parameter is expected to be determined at 90\% CL within 3 years of operation of the full ORCA detector~\cite{KM3NeT:2023ncz}, leading to a stronger constraint on $\alpha_3$. 

\begin{figure}[]
 \centering
     \includegraphics[width=0.75\linewidth]{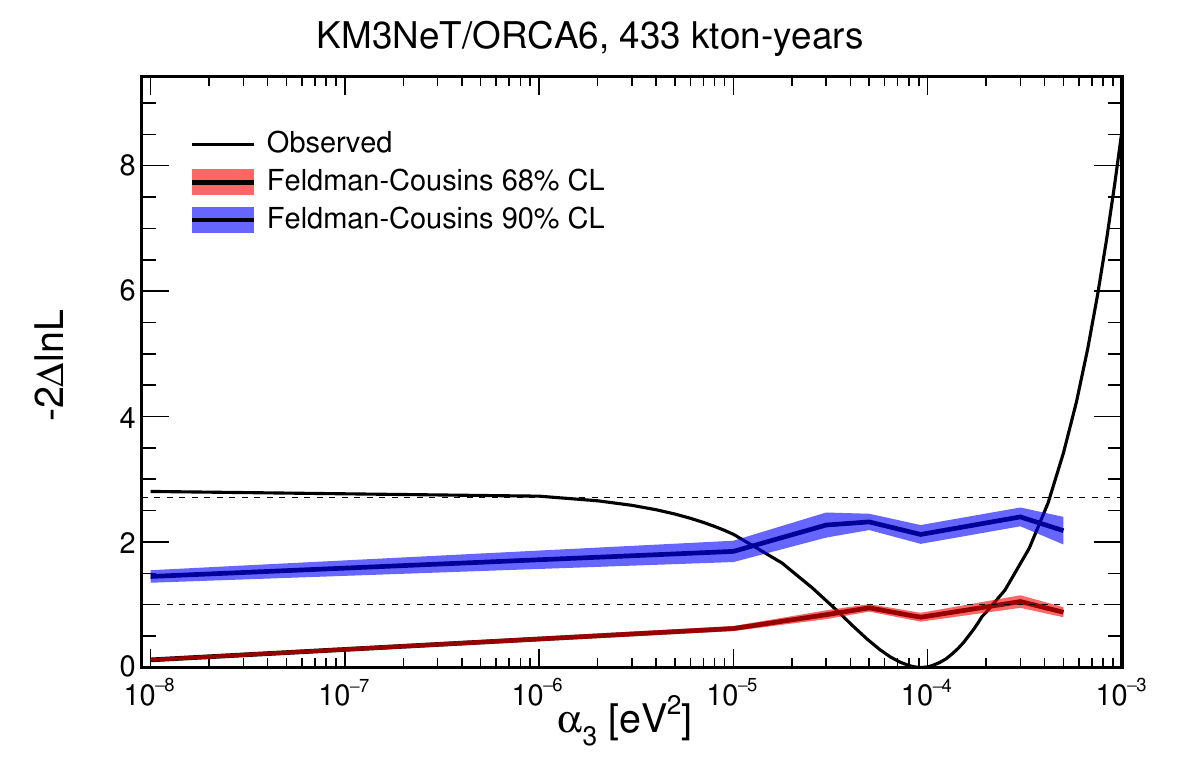} 
    \caption{Profiled log-likelihood ratio of the invisible decay parameter, $\alpha_{3}$. The black line represents the observed result. 
Horizontal dashed lines represent the 68\% and 90\% CL thresholds assuming Wilks' theorem, while the red and blue lines show respectively the 68\% and 90\% Feldman-Cousins CL. The uncertainty bands are the standard deviation with respect to the CLs derived by sampling the pseudo-experiments with replacement.
    }
    \label{Fig:DataBandsD}
\end{figure}

In order to compute Feldman-Cousins corrections to the parameter uncertainties, a set of 2000 pseudo-experiments is generated for several testing points along the profile. The results compared to the expectation from Wilks’ theorem
can be seen in figure \ref{Fig:DataBandsD}. The decrease of Feldman-Cousins threshold values for $\alpha_3 \rightarrow 0$ reflects the fact that the validity conditions of Wilks' theorem are not met due to the presence of boundaries in the parameter space \cite{article}.

\begin{figure}[]
    \centering
    \includegraphics[width=0.6\linewidth]{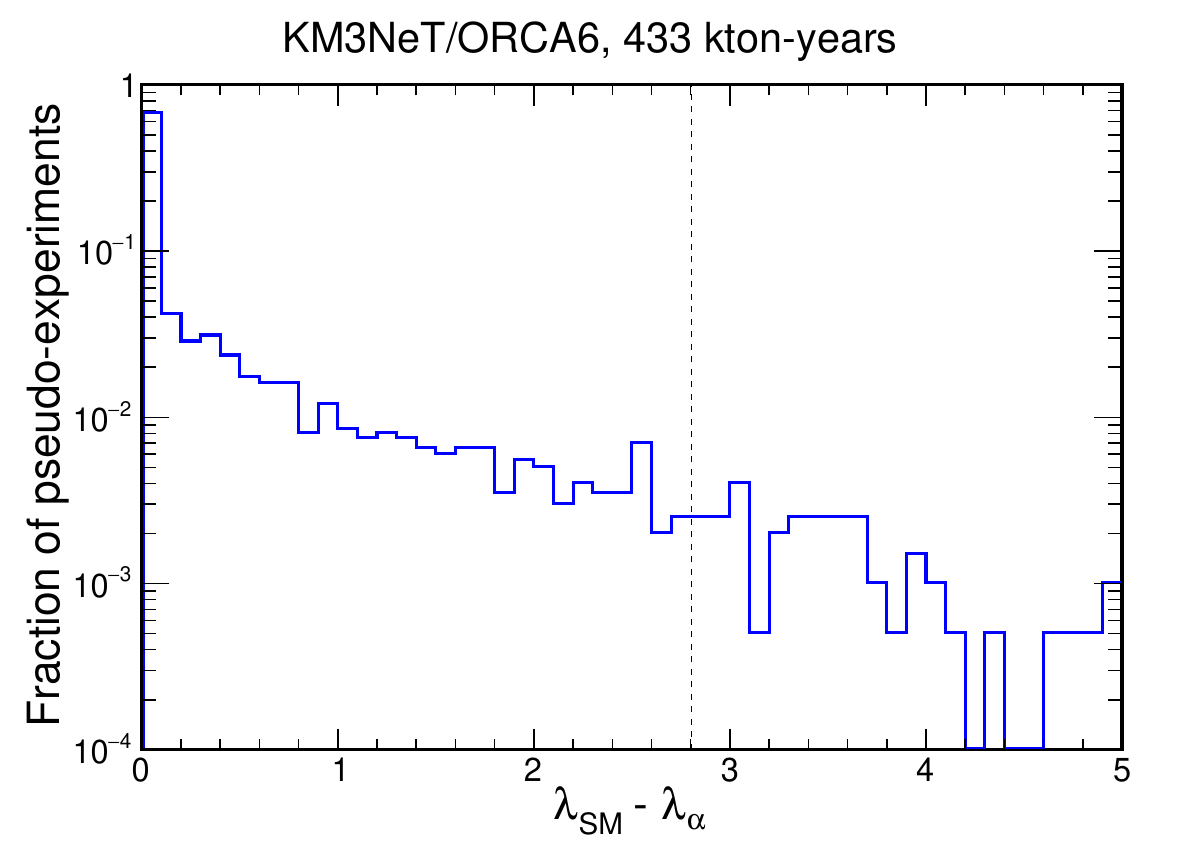}
    \caption{Distribution of the test statistic $\Delta \lambda$, obtained from a set of 2000 pseudo-experiments generated assuming as true NuFit v5.0 values and $\alpha_3=0$. The vertical line indicates the observed test statistic value.}
    \label{Fig:C6Decay_Hyp}
\end{figure}

The difference in the likelihood of the best fit to the standard oscillation hypothesis is $\Delta \lambda_{\text{obs}}=2.8$. The corresponding p-value is computed by simulating pseudo-experiments using as true parameter values the NuFit v5.0 best fit and computing the log-likelihood ratio $\Delta \lambda = - 2 (\log L_{SM} - \log L_D)$ between the Standard Model (SM) hypothesis ($\alpha_3=0$) and the invisible neutrino decay (D) hypothesis ($\alpha_3$ free). The distribution of the log-likelihood ratio is shown in figure \ref{Fig:C6Decay_Hyp}, together with the value observed from ORCA6 data analysis. The probability of getting $\Delta \lambda$ equal or higher than $\Delta \lambda_{\text{obs}}$ is $(3.9 \pm $0.4$)\%$, corresponding to $2.1\,\sigma$.

\begin{figure}[]
 \centering
   \includegraphics[width=0.75\linewidth]{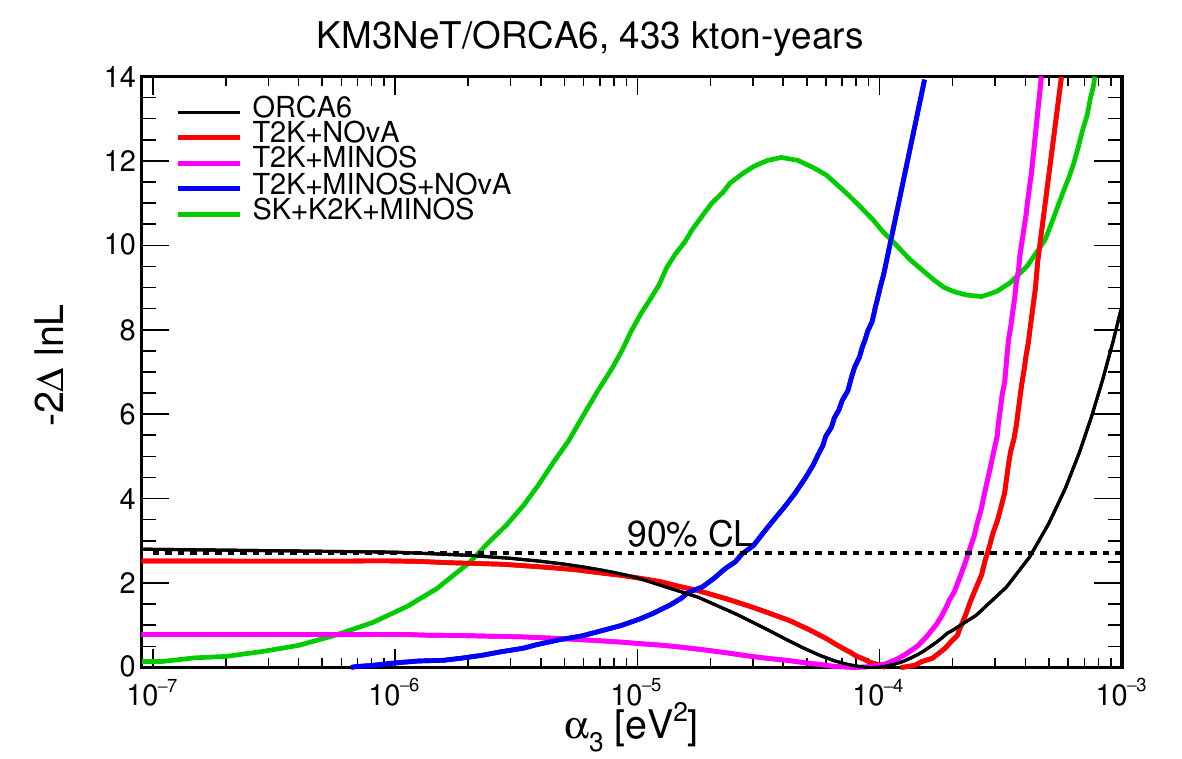}
 \caption{The observed profiled log-likelihood scan (black) compared with combined fits using data from: T2K+NO$\nu$A (red) \cite{NovaT2K}, T2K+MINOS (magenta) \cite{MinosT2K}, T2K+MINOS+NO$\nu$A (blue) \cite{MINOST2KNOVA} and SK+K2K+MINOS (green) \cite{SKK2KMINOS}.}
 \label{C6-DFC}
\end{figure}

In figure \ref{C6-DFC}, the ORCA6 result is compared with the ones obtained from combined fits using long-baseline and atmospheric neutrino data (see table \ref{Table5_AllExp}). The data best-fit and intervals are in the same order of magnitude as those of combined fits with long baseline data, T2K+NO$\nu$A (red) \cite{NovaT2K} and T2K+MINOS (magenta) \cite{MinosT2K} and one order of magnitude weaker than the combination of T2K+MINOS+NO$\nu$A (blue) \cite{MINOST2KNOVA}. Note that for the combination of SK, K2K and MINOS data (green) \cite{SKK2KMINOS}, a two-flavour approximation is applied and matter effects are neglected, so that a complete treatment may relax the second minimum. Future accelerator experiments, such as DUNE \cite{DUNEupdated}, MOMENT \cite{MOMENT_Tang_2019}, ESSnuSB \cite{ESSnuSB_choubey2020exploring} and T2HKK \cite{Dey_2024}, as well as reactor experiments like JUNO \cite{abrahao2015constraint}, and atmospheric neutrino experiments like INO \cite{Choubey_2018_INO}, are expected to provide an improved sensitivity to this parameter.

\begin{table}[]
    \centering
    \begin{tabular}{|l|r|c|}
    \hline
    Experiment   &   UL ($90\%$ CL) [$\mathrm{10^{-6} eV^2}$] &  Reference\\
    \hline
    \textbf{ORCA6 (433 kton-year)} &  [10, 380]   & this paper \\ \hline
    \textbf{ORCA (70 Mton-year)} & 3.7  & \cite{KM3NeT:2023ncz}\\ \hline
    \textcolor{blue}{T2K, NO$\nu$A} &\textcolor{blue}{290}  &   \cite{NovaT2K} \\ \hline
    \textcolor{blue}{T2K, MINOS} &\textcolor{blue}{240}  &   \cite{MinosT2K} \\ \hline
    \textcolor{blue}{T2K, NO$\nu$A, MINOS} &\textcolor{blue}{27}  &   \cite{MINOST2KNOVA} \\ \hline
    \textcolor{blue}{K2K, MINOS, SK I+II} &\textcolor{blue}{2.3}  &  \cite{SKK2KMINOS}\\ \hline
    DUNE (5$\nu$+5$\bar{\nu}$) yr &13  &  \cite{DUNEupdated} \\ \hline
    MOMENT (10 yr) & 24  &\cite{MOMENT_Tang_2019} \\ \hline
    ESSnuSB (5$\nu$+5$\bar{\nu}$) yr &$16-13$  & \cite{ESSnuSB_choubey2020exploring} \\  \hline
    
    JUNO (5 yr) & 7 &  \cite{abrahao2015constraint} \\  \hline
    INO-ICAL (10 yr) & 4.4& \cite{Choubey_2018_INO} \\ \hline
    \end{tabular}
    \caption{Average upper limits (ULs) at $90\%$ CL for the decay parameter $\alpha_3$ for combined fits (blue) and future experiments. 70 Mton-years of ORCA correspond to 10 years of data taking with the full detector. This ORCA6 analysis provides an interval at 90\% CL after Feldman-Cousins corrections; an upper limit can be set above 96\% CL.} 
  \label{Table5_AllExp}
\end{table}

\begin{figure}[]
\centering

      \includegraphics[width=0.45\linewidth]{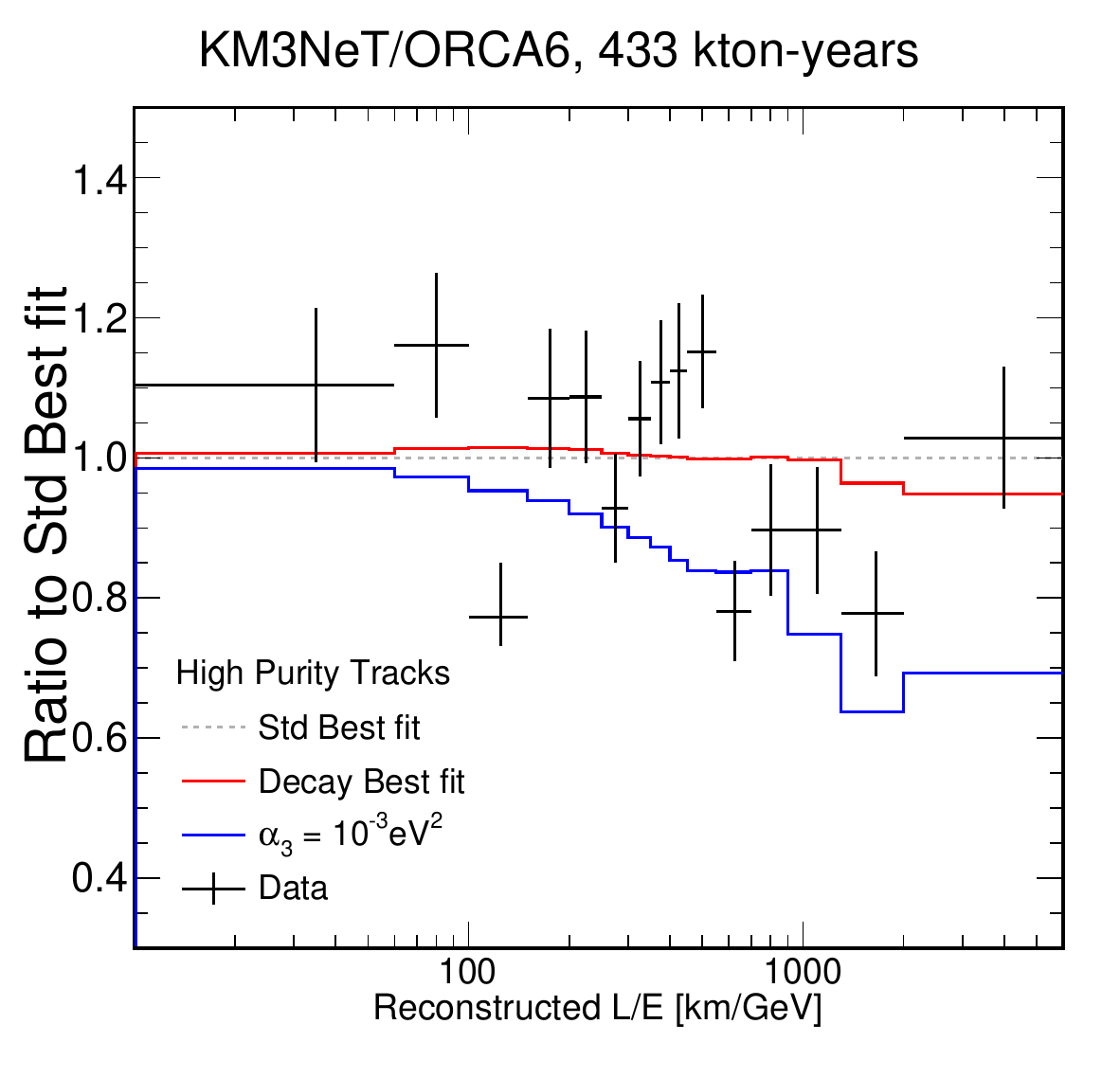}
      \includegraphics[width=0.45\linewidth]{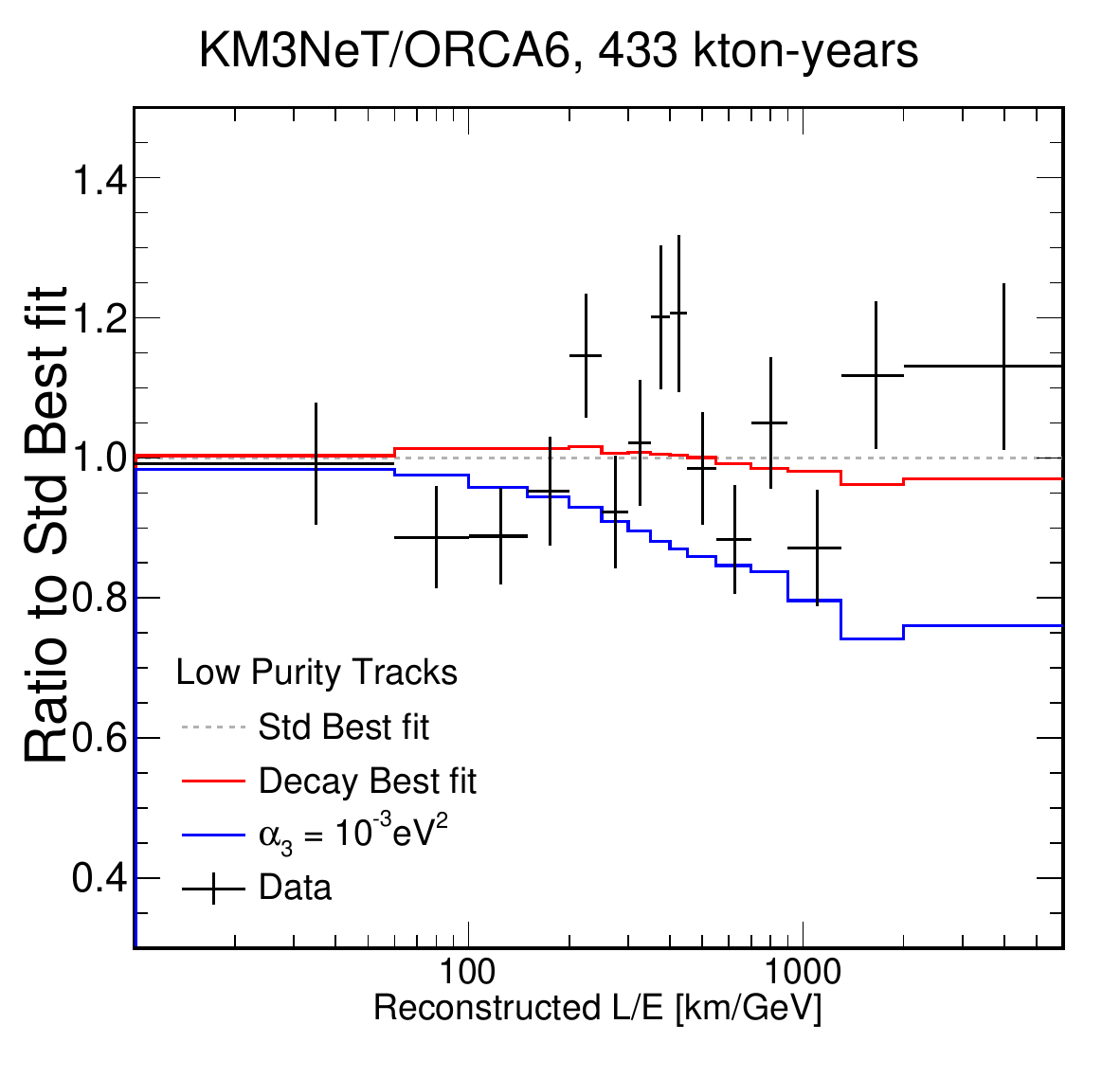} \\
      \includegraphics[width=0.45\linewidth]{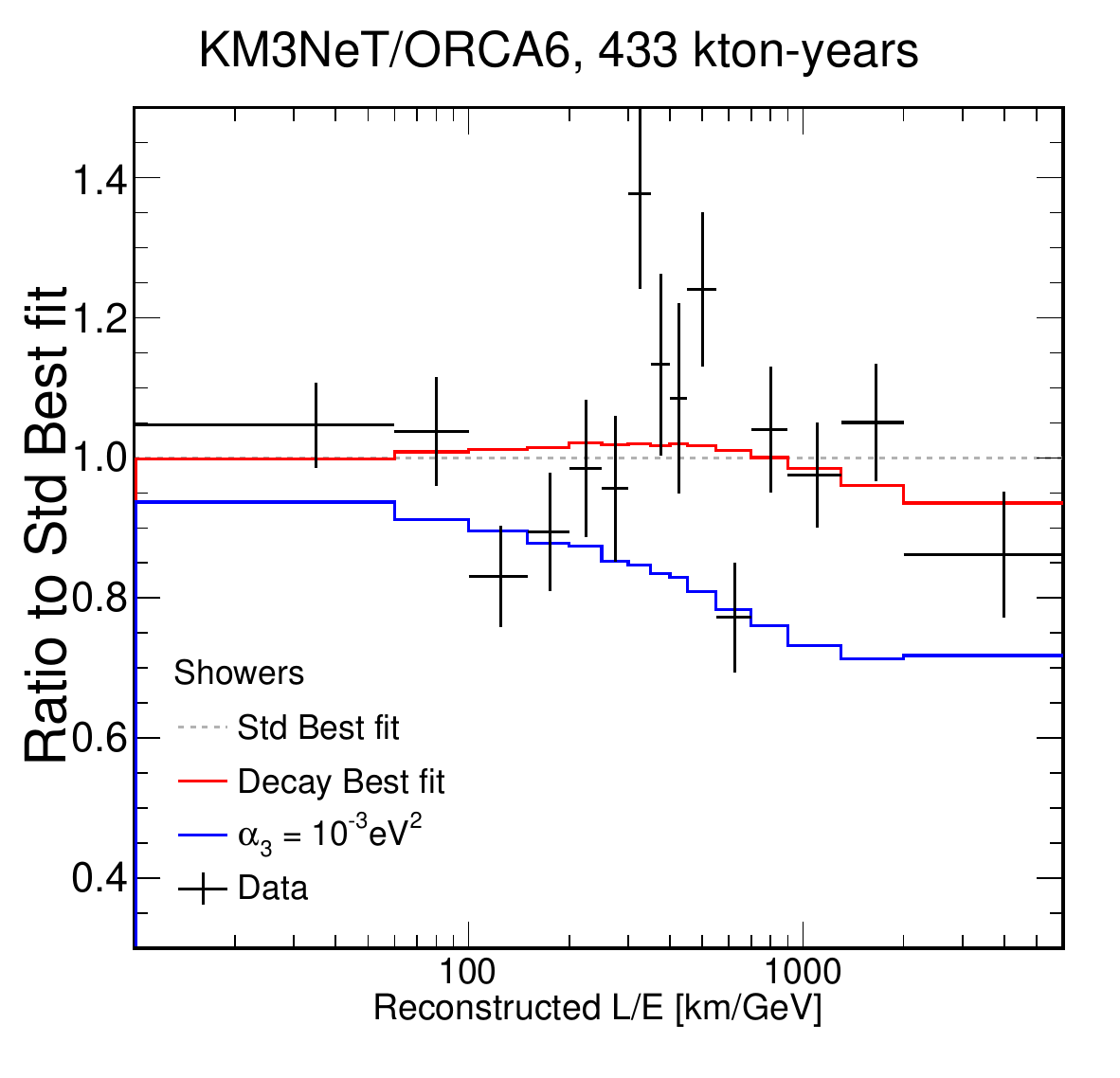}
 \caption{Ratio to the standard oscillations best fit of the reconstructed length over energy ratio, $L/E$, for High Purity Tracks (upper-left), Low Purity Tracks (upper-right) and Showers (bottom). The data points are shown with error bars in black, the best fit for standard oscillation hypothesis in dashed grey, the best fit assuming invisible neutrino decay in solid red, and an extreme case of decay ($\alpha_3=1.1\times 10^{-3}~\mathrm{eV^2}$) in solid blue. }
 \label{Fig:LoED}
\end{figure}

The ORCA6 event distributions for each of the three event classes are presented in figure~\ref{Fig:LoED} as a function of the reconstructed $L/E$ ratio. In each plot, the data are compared to the standard oscillation best fit, the invisible decay best fit and to an extreme decay case ($\alpha_3=1.1\times 10^{-3}~\mathrm{eV^2}$), using the nuisance parameters the best-fit values to the decay hypothesis. The underfluctuation at large $L/E$ values in the High Purity Track class is compatible with the decay hypothesis, but no such clear decrease is visible in the two other event classes.

\clearpage

\begin{figure}[H]
 \centering
     \includegraphics[width=0.69\linewidth]{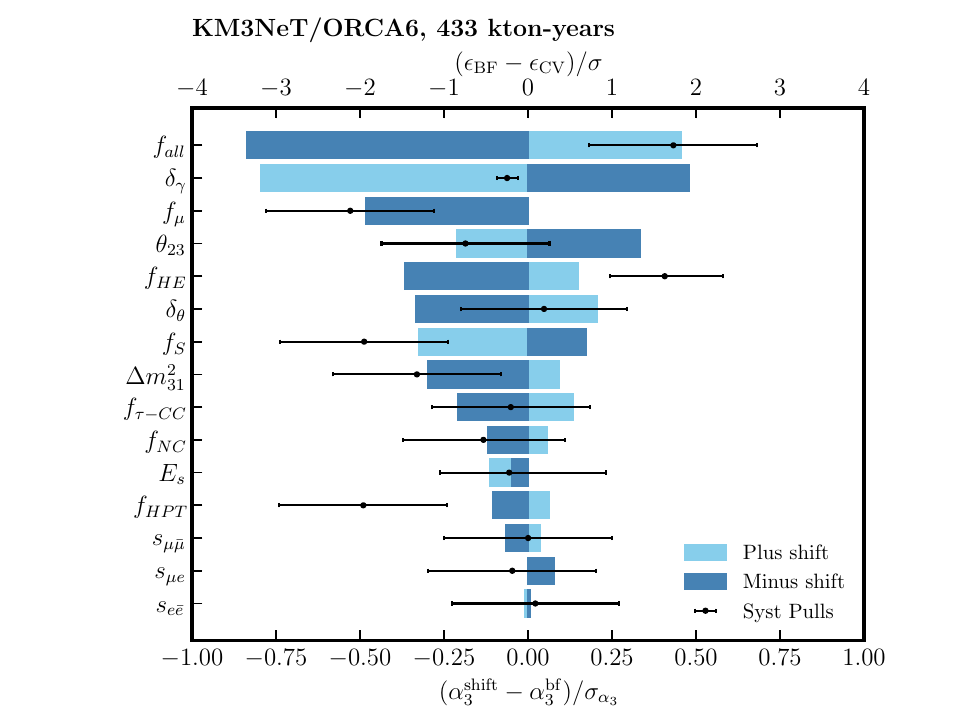}
 \caption{Impact of the nuisance parameters on $\alpha_{3}$  evaluated by
repeating the fit shifting the nuisance parameters by their post-fit uncertainties and comparing the fitted value $\alpha_3^{\text{shift}}$ to the best-fit value, $\alpha_3^{bf}$ (bottom axis). The pulls of the nuisance parameters
are reported as dots (top axis) with bars representing the ratio between the post-fit uncertainties and the pre-fit uncertainties (the ratio is set to 1 for unconstrained nuisance parameters). }
 \label{Fig:DataPullsD}
\end{figure}

The impact of each nuisance parameter on the estimation of the parameters of interest is evaluated separately, by repeating the fit while shifting the parameter value up and down by its own post-fit uncertainty. Then, the
overall best-fit value of the parameter of interest is compared with the fit obtained with the
"shifted" values.
The difference between the nominal best fit of the parameter of interest and the
"shifted" value normalised to its 1~$\sigma$ uncertainty is reported with boxes in figure~\ref{Fig:DataPullsD}. Additionally, the pulls of the best-fit (BF) nuisance parameter values with
respect to the central values (CV), ($\epsilon_{\text{BF}}-\epsilon_{\text{CV}})/\sigma$, are reported as dots with error bars where $\sigma$
represents their pre-fit uncertainty. Error bars for the pulls are defined as the ratio between post-fit and pre-fit
uncertainties. In the case of unconstrained parameters, where no pre-fit uncertainty is set, the post-fit uncertainty is used instead.

The invisible decay parameter is mostly affected by the flux normalisation factor, the spectral index $\delta_{\gamma}$ and the horizontal-to-vertical ratio $\delta_{\theta}$ (as the events most affected by invisible decay come from low energies and longest paths), and $\theta_{23}$. The spectral index and the normalisation of high energy simulated events can be constrained better with the data than with the auxiliary measurements, as can be seen from the small bars. The pull in the normalisation of high-energy simulated events is expected, because
the light simulation software used for high energies leads to a reduced number of selected events with respect to the simulation made with KM3Sim.


\section{Conclusions}

Based on a sample of 5828 neutrino candidates collected with the ORCA6 detector, the invisible neutrino decay parameter $\alpha_3$ has been measured. The result is:

\begin{align*}
&\alpha_3= 0.92^{+1.08}_{-0.57}\times 10^{-4} ~\mathrm{eV^2}.
\end{align*}
The corresponding likelihood ratio value is $2\ln(L_{\alpha}/L_{\text{SM}}) = 2.8 $ and the associated p-value is $(3.9 \pm 0.4)\%$,  
corresponding to $2.1\,\sigma$ compatibility with the Standard Model hypothesis of no neutrino decay.

The constraint on the invisible decay parameter obtained in this study is compatible with results from combined fits with data from long-baseline neutrino experiments and is of the same order of magnitude. This result indicates that even at early stages of KM3NeT/ORCA, with a small detector and limited statistics, the potential to probe scenarios beyond the Standard Model is significant.

\section{Acknowledgements}

The authors acknowledge the financial support of:
KM3NeT-INFRADEV2 project, funded by the European Union Horizon Europe Research and Innovation Programme under grant agreement No 101079679;
Funds for Scientific Research (FRS-FNRS), Francqui foundation, BAEF foundation.
Czech Science Foundation (GAČR 24-12702S);
Agence Nationale de la Recherche (contract ANR-15-CE31-0020), Centre National de la Recherche Scientifique (CNRS), Commission Europ\'eenne (FEDER fund and Marie Curie Program), LabEx UnivEarthS (ANR-10-LABX-0023 and ANR-18-IDEX-0001), Paris \^Ile-de-France Region, Normandy Region (Alpha, Blue-waves and Neptune), France,
The Provence-Alpes-Côte d'Azur Delegation for Research and Innovation (DRARI), the Provence-Alpes-Côte d'Azur region, the Bouches-du-Rhône Departmental Council, the Metropolis of Aix-Marseille Provence and the City of Marseille through the CPER 2021-2027 NEUMED project,
The CNRS Institut National de Physique Nucléaire et de Physique des Particules (IN2P3);
Shota Rustaveli National Science Foundation of Georgia (SRNSFG, FR-22-13708), Georgia;
This work is part of the MuSES project which has received funding from the European Research Council (ERC) under the European Union’s Horizon 2020 Research and Innovation Programme (grant agreement No 101142396).
This work was supported by the European Research Council, ERC Starting grant \emph{MessMapp}, under contract no. $949555$.
The General Secretariat of Research and Innovation (GSRI), Greece;
Istituto Nazionale di Fisica Nucleare (INFN) and Ministero dell’Universit{\`a} e della Ricerca (MUR), through PRIN 2022 program (Grant PANTHEON 2022E2J4RK, Next Generation EU) and PON R\&I program (Avviso n. 424 del 28 febbraio 2018, Progetto PACK-PIR01 00021), Italy; IDMAR project Po-Fesr Sicilian Region az. 1.5.1; A. De Benedittis, W. Idrissi Ibnsalih, M. Bendahman, A. Nayerhoda, G. Papalashvili, I. C. Rea, A. Simonelli have been supported by the Italian Ministero dell'Universit{\`a} e della Ricerca (MUR), Progetto CIR01 00021 (Avviso n. 2595 del 24 dicembre 2019); KM3NeT4RR MUR Project National Recovery and Resilience Plan (NRRP), Mission 4 Component 2 Investment 3.1, Funded by the European Union – NextGenerationEU,CUP I57G21000040001, Concession Decree MUR No. n. Prot. 123 del 21/06/2022;
Ministry of Higher Education, Scientific Research and Innovation, Morocco, and the Arab Fund for Economic and Social Development, Kuwait;
Nederlandse organisatie voor Wetenschappelijk Onderzoek (NWO), the Netherlands;
The grant “AstroCeNT: Particle Astrophysics Science and Technology Centre”, carried out within the International Research Agendas programme of the Foundation for Polish Science financed by the European Union under the European Regional Development Fund; The program: “Excellence initiative-research university” for the AGH University in Krakow; The ARTIQ project: UMO-2021/01/2/ST6/00004 and ARTIQ/0004/2021;
Ministry of Research, Innovation and Digitalisation, Romania;
Slovak Research and Development Agency under Contract No. APVV-22-0413; Ministry of Education, Research, Development and Youth of the Slovak Republic;
MCIN for PID2021-124591NB-C41, -C42, -C43 and PDC2023-145913-I00 funded by MCIN/AEI/10.13039/501100011033 and by “ERDF A way of making Europe”, for ASFAE/2022/014 and ASFAE/2022 /023 with funding from the EU NextGenerationEU (PRTR-C17.I01) and Generalitat Valenciana, for Grant AST22\_6.2 with funding from Consejer\'{\i}a de Universidad, Investigaci\'on e Innovaci\'on and Gobierno de Espa\~na and European Union - NextGenerationEU, for CSIC-INFRA23013 and for CNS2023-144099, Generalitat Valenciana for CIDEGENT/2018/034, /2019/043, /2020/049, /2021/23, for CIDEIG/2023/20, for CIPROM/2023/51 and for GRISOLIAP/2021/192 and EU for MSC/101025085, Spain;
Khalifa University internal grants (ESIG-2023-008, RIG-2023-070 and RIG-2024-047), United Arab Emirates;
The European Union's Horizon 2020 Research and Innovation Programme (ChETEC-INFRA - Project no. 101008324).

\newpage
\bibliographystyle{JHEP}
\bibliography{main}
\end{document}